\documentclass[useAMS,usenatbib]{mn2e} \newif\ifdraft \draftfalse
\usepackage{graphics}
\usepackage{rotating}
\usepackage{amsmath}
\usepackage{amssymb}
\usepackage{color}
\usepackage{hyperref}
\hypersetup{
            colorlinks=true,
            linkcolor=black,
            filecolor=black,
            citecolor=black,
            urlcolor=blue
           }

\renewcommand{\aa}{A\&A}
\newcommand{\aj}{AJ}
\newcommand{\apj}{ApJ}
\newcommand{\apjs}{ApJS}

\newcommand{\mnras}{MNRAS}

\renewcommand{\fs}{\footnotesize}

\newcommand{\ns}{\normalsize}

\newcommand{\getlength}[1]{\ifx#1\end \let\next=\relax
            \else\advance\count255 by1 \let\next=\getlength\fi \next}
\newcommand{\ifnularg}[1]{ \count255=0 \getlength#1\end \ifnum\count255=0 }
\newcommand{\ifm}{\makebox{}\ifmmode}
\long\def\ifundefined#1#2#3{\expandafter\ifx\csname
  #1\endcsname\relax#2\else#3\fi}
\newcommand{\beq}   { \begin{eqnarray} }
\newcommand{\eeq}[1]{ \ifnularg{#1} end{eanarray} \else
                      \label{#1}\end{eqnarray}    \fi }
\newcommand{\eeql}   { \end{eqnarray} }
\newcommand{\eeqn}   { \nonumber \end{eqnarray} }
\newcommand{\Frac}[2]{\frac{\displaystyle\strut #1}{\displaystyle\strut #2} }

\newcommand{\dss}{\displaystyle}

\newcommand{\ntab}[2]{ \multicolumn{1}{#1}{#2} }

\newcommand{\hem}{\hspace{1em}}
\newcommand{\hhem}{\hspace{0.5em}}

\newcommand{\Number}[1]{\ifnum#1<10\relax0\number#1\else\number#1\fi}
\newcommand{\isodate}{
\count151=\time
\divide\count151 by 60
\count151=\count151
\multiply\count151 by 60
\count152=\time
\advance\count152 by -\count151
\divide\count151 by 60
\count152=\count151
\multiply\count151 by 60
\count153=\time
\advance\count153 by -\count151
\Number{\year}.\Number{\month}.\Number{\day}--\Number{\count152}:\Number{\count153}
}
\definecolor{Dred}{rgb}{0.312,0.070,0.070}
\definecolor{Dblue}{rgb}{0.070,0.070,0.312}
\definecolor{Dgreen}{rgb}{0.070,0.312,0.070}
\definecolor{Db}{rgb}    {0.050,0.0,0.320}

\newcommand{\Blb}[1]{\textcolor{Dblue}{\bf #1}}

\newcounter{note}
\let\oldmarginpar\marginpar
\renewcommand\marginpar[1]{\-\oldmarginpar[\raggedleft\footnotesize #1]%
{\raggedright\footnotesize #1}}
\newcommand{\note}[1]{{\color{Dred}\bf #1}}
\newcommand{\Note}[1]{\Rdb{#1}{\addtocounter{note}{1}%
\marginpar{\small\underline{\Rdb{Corr \arabic{note}}}}}}
\renewcommand{\note}[1]{#1}
\renewcommand{\Note}[1]{#1}
     \newcommand{\web}[1]{\Blb{\url{#1}}}
\newcommand{\Fermi}{{\it Fermi}}
\newcommand{\task}[1]{\textsl{#1}}

\ifdraft \voffset=-4mm \fi  

\volume{}
\pubyear{}
\pagerange{}

\newcommand{\published}{Published 2013 April 28; Accepted 2013 MArch 26;
                      Received 2013 March 9; in original form 2013 January 10}

\title[
ATCA observations of Fermi unassociated sources
]{
Australia Telescope Compact Array observations of Fermi unassociated sources
}
\author[Petrov et al.]{
  \parbox[t]{\textwidth}{
     Leonid Petrov$^{1}$\thanks{E-mail:Leonid.Petrov@lpetrov.net},
     Elizabeth K. Mahony$^{2}$,
     Philip G. Edwards$^{3}$,
     \mbox{Elaine M.~Sadler}$^{4}$,
     \mbox{Frank K. Schinzel}$^{5}$,
     \mbox{David McConnell}$^{3}$
  }
\vspace{1.0ex} \\
$^{1}$Astrogeo Center, Falls Church, VA 22043 USA \\
$^{2}$ASTRON, the Netherlands Institute for Radio Astronomy, Postbus 2, 7990 AA, Dwingeloo, The Netherlands \\
$^{3}$CSIRO Astronomy and Space Science, PO Box 76, Epping, NSW 1710,
      Australia \\
$^{4}$Sydney Institute for Astronomy, School of Physics, The University of Sydney, NSW 2006, Australia \\
$^{5}$Department of Physics and Astronomy, University of New Mexico, Albuquerque NM, 87131, USA \\
  \phantom{A} \\ \phantom{A} \\ 
\published
}

\pagerange{\pageref{firstpage}--\pageref{lastpage}}

\begin{document}

\maketitle
\label{firstpage}

\begin{abstract}

\ifdraft
\fi


   We report results of the first phase of observations with the Australia 
Telescope Compact Array (ATCA) at 5 and 9~GHz of the fields around 411 
$\gamma$-ray sources with declinations $<+10\degr$ detected by
\Fermi\ but marked as unassociated in the 2FGL catalogue. \note{We have 
detected 424 sources with flux densities in a range of 
2~mJy to 6~Jy that lie within the 99 per cent localisation uncertainty of 283 
$\gamma$-ray sources. Of these, 146 objects were detected in 
both the 5 and 9~GHz bands. We found 84 sources in our sample with a spectral index flatter
than -0.5. The majority of detected sources are weaker than 
100 mJy and for this reason were not found in previous surveys. 
Approximately 1/3 of our sample, 128 objects, have the probability of being 
associated by more than 10 times than the probability of being a background 
source found in the vicinity of a $\gamma$-ray object by chance.} We present 
the catalogue of positions of these sources, estimates of their flux 
densities and spectral indices where available.

\end{abstract}


\section{Introduction}

  Analysis of the first two years of \Fermi/LAT observations yielded a catalogue 
of 1872 $\gamma$-ray sources \citep[][hereafter 2FGL]{r:2fgl}. 
Of these, 70 per cent have associations with blazars, pulsars, supernova 
remnants or other objects. The procedure for assigning associations
is described in full detail in \citet{r:fermi_agn}. However, the 2FGL
catalogue still does not provide associations for 573 $\gamma$-ray sources. 
The positional accuracy of \Fermi\ ranges from 0.1~arcmin to 16.6~arcmin 
with a median $1\sigma$ uncertainty of 2.0~arcmin. These large position 
errors prevent finding associations by direct matching the 2FGL 
against optical or infrared catalogues.

  Previous analysis of the 1FGL catalogue~\citep{r:Fermi_VCS} has confirmed 
earlier EGRET results: that $\gamma$-ray emission and parsec-scale radio 
emission are strongly related. Extending this study to the 2FGL
catalogue, we found that 770 out of 1872 \Fermi\ sources, roughly 
one half, have been detected in VLBI surveys at 8~GHz (as of December 2012).
All these objects are Active Galactic Nuclei (AGN). This is the dominant
population of point-like \Fermi\ sources outside of the Galactic plane. 
Other types of objects associated with $\gamma$-ray sources are supernova 
remnants, novae, pulsar wind nebulae, x-ray binaries, microquasars, 
and pulsars. Unassociated sources may belong to any of these classes 
or may, in part, constitute an as-yet-unknown population. Due to the 
correlation of $\gamma$-ray emission with parsec-scale radio emission, 
high resolution radio observations are useful for classification 
of unassociated $\gamma$-ray sources. Supernova remnants and pulsar wind 
nebulae are extended objects and, as such, high resolution radio 
interferometric observations tend to resolve out their emission making 
them undetectable. Pulsars are generally weak at high frequencies 
($>5$~GHz) as they generally have steep spectral indices. Therefore, 
sources detected with VLBI that are brighter than 1~mJy are almost always 
AGNs. Detection of a radio bright AGN within the \Fermi\ position error 
ellipse presents a strong argument that this is likely the same object. 
Therefore, systematic high resolution radio observations of unassociated 
\Fermi\ objects promise to find all $\gamma$-ray sources associated with 
radio-loud AGNs and shrink the list of objects that remain unassociated 
with an astrophysical source.

  It was shown by \citet{r:Lister09} and \citet{r:Fermi_VCS} that 
$\gamma$-ray fluxes and 8~GHz radio flux densities from regions 
smaller than 5~mas are correlated. Source variability confirms 
that both radio and $\gamma$-ray emission comes from parsec-scale regions.
Therefore, a flux-limited sample of $\gamma$-ray AGNs, 
such as 2FGL, should be contained within a flux-limited sample of 
compact radio sources. Compact radio sources from a flux-limited 
catalogue found within the position error ellipses of sources from 
a $\gamma$-ray flux limited catalogue can be considered with a high 
confidence as the same objects, provided that the probability of finding 
a background radio-loud source within the error ellipse is small. 
As the search area becomes smaller, weaker radio-loud AGNs can be 
associated with a $\gamma$-ray source.

\begin{figure}
   \includegraphics[width=0.48\textwidth]{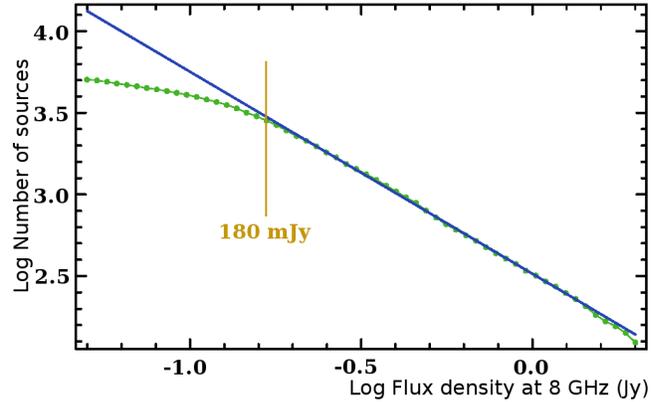}
   \caption{Dependence of the decimal logarithm of the number of 
            compact sources versus the decimal logarithm of their
            minimum correlated flux density from regions smaller 
            than 50~mas at 8~GHz.}
   \label{f:logn_logs}
\end{figure}

  In order to evaluate the probability of finding a background source
in a given search area, we investigated the cumulative all-sky catalogue 
of compact radio sources detected with VLBI in the absolute astrometry mode 
(Petrov and Kovalev, in preparation). This catalogue\footnote{Available
at \web{http://astrogeo.org/rfc}}, as of December 2012, had 7215 objects 
detected in numerous VLBI surveys over the last several decades: Very Long 
Baseline Array (VLBA) Calibrator Survey 
\citep{r:vcs1,r:vcs2,r:vcs3,r:vcs4,r:vcs5,r:vcs6}, 
Long Baseline Array (LBA) calibrator survey \citep{r:lcs1}, VLBA Galactic 
Plane Survey \citep{r:vgaps}, European VLBI Network (EVN) Galactic plane 
survey \citep{r:egaps}, VLBA imaging and polarimetry 
survey \citep{r:tay07,r:vips}, regular VLBA geodetic observations 
\citep{r:rdv,r:pus12}, and ongoing VLBI observations of 2MASS 
galaxies \citep{r:v2m}. Figure~\ref{f:logn_logs} shows the dependence 
of the logarithm of the number of sources in the whole sky $N$ with a flux 
density greater than $S$, as a function of the logarithm of $S$ determined 
as the median correlated flux density at baseline projection lengths 
in a range of 100--900~km at 8~GHz. We see that the dependence can 
be approximated as $ N(S) \approx 327 \cdot S^{-1.237}$ in a range of 
[0.18, 5]~Jy. We interpret the deviation of $\log N(\log S)$ from the 
straight line below 180~mJy as evidence of incompleteness. Assuming the 
parent population remains the same for sources as weak as 1~mJy, we can 
extrapolate the number of compact sources 
in the celestial sphere derived for the range [0.18, 5]~Jy to 
the range [1, 180]~mJy. 
%

  Let us consider a search for a radio counterpart within the area 
where the probability of source localisation is $P$. Then, assuming
localisation errors follow the 2D normal distribution with
the second moment $\sigma$, the mathematical expectation of 
the number of background sources that can be found in that area is

\beq
    M(\sigma,S,P) = -\Frac{\sigma^2}{2\pi}\, \ln(1-P)\, N(S).
\eeq{e:e2}

\begin{figure}
   \includegraphics[width=0.48\textwidth]{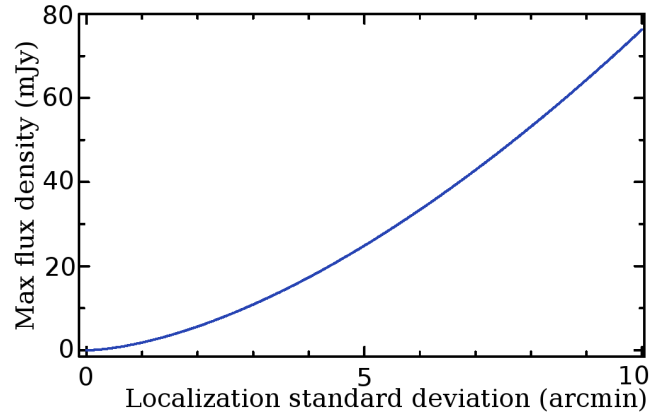}
   \caption{Maximum flux density at 8~GHz of a background compact source
            to be found with the probability of 10 per cent within an error 
            ellipse with a given $1\sigma$ standard deviation.
            }
   \label{f:prob_err}
\end{figure}

  The function $M(\sigma,S)$ depends on both the position uncertainty,
the flux density, and the probability of localisation. When 
$M \ll 1$, it can be interpreted as the probability of at least one 
source being found within the area of position uncertainty. If we 
fix $M(\sigma,S)$ to a specific value, for instance, 0.1, and 
fix localisation probability, e.g. to 0.95, we can find the dependence 
of the maximum flux density $S$ of a background source that can be
found within the area of localisation with a certain probability
(10~per cent in our example) on the standard deviation of localisation.
This dependence is shown in Figure~\ref{f:prob_err}. For the median 
1$\sigma$ position uncertainty of unassociated sources from the 2FGL 
catalogue 3.0~arcmin, the probability to find a background source brighter 
than 10.9~mJy is 10~per cent. For 80~per cent of 2FGL source, $1\sigma$ 
position uncertainty is less than 4.0~arcmin. The flux density that 
corresponds to 10~per cent probability to find a background source with 
such a position uncertainty is 17.3~mJy.

\begin{figure}
   \includegraphics[width=0.48\textwidth]{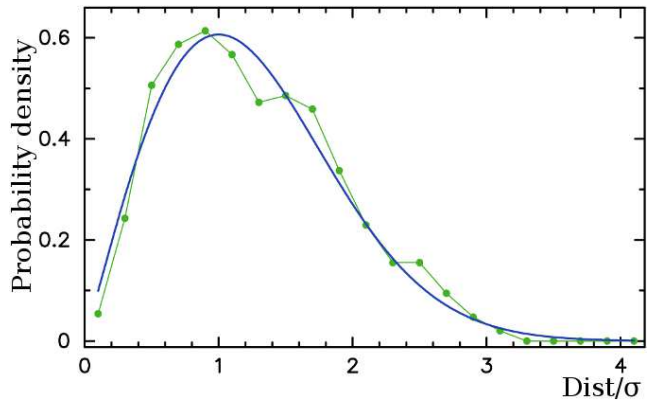}
   \caption{Distribution of position offsets normalised by their
            standard deviations between 770 \Fermi\ sources and their 
            counterparts detected with VLBI. The solid blue lines shows
            Rayleigh distribution with $\sigma=1$.
            }
   \label{f:distr}
\end{figure}

   In order to check the validity of the reported 2FGL position uncertainties,
we computed arcs between 770 $\gamma$-ray objects that have counterparts
with radio sources observed with VLBI, and normalised them to their standard
deviations derived from parameters of their 2FGL error ellipses:

\beq
  n = \sqrt{\Delta^2\alpha \cos^2 \delta + \Delta^2\delta} \:
      \Frac{\sqrt{\sigma^2_{\rm maj} \sin^2 \beta + \sigma^2_{\rm min} \cos^2 \beta}}
           {\sigma_{\rm maj} \, \sigma_{\rm min} },
\eeq{e:e3}
   where angle $\beta$ is 
\beq
  \beta = \arctan{\Frac{\Delta\delta}{\Delta\alpha \, \cos\delta} - (\pi/2 - \theta)}
\eeq{e:e4}
   and $\sigma_{\rm maj}$, $\sigma_{\rm min}$, and $\theta$ are
semi-major, semi-minor axes and position angle of the 2D Normal
distribution that describes errors of source localisation. We derived
$\sigma_{\rm maj}$, $\sigma_{\rm min}$ from the so-called ``95\% 
confidence source semi-major and semi-minor axes'' reported in table~3
of the 2FGL catalogue by scaling them by 1/e. The distribution of
normalised arcs between the VLBI and 2FGL positions follows a Rayleigh 
distribution with $\sigma=1$ very closely (Figure~\ref{f:distr}). 
The histogram is best fit into a Rayleigh distribution with 
$\sigma = 1.026$. Since VLBI positions are five orders of magnitude 
more accurate than the positions reported in 2FGL, they are considered 
as true. We conclude that the 2FGL position errors are realistic and 
can be used for a probabilistic inference.

   These estimates show that radio observations are an efficient way
to identify AGNs associated with $\gamma$-ray sources. 
\Note{According to our experience, it is not difficult to detect 
sources with flux densities of several mJy using radio interferometers.} 
As we have shown in Figure~\ref{f:logn_logs}, the cumulative catalogue 
of compact radio sources detected with VLBI is complete only to the level 
of 180~mJy. Sources with correlated flux densities between 2 and 180~mJy may 
not have been marked as associations with 2FGL objects because they were 
missing from VLBI surveys. However, not all associations of $\gamma$-ray 
sources in 2FGL are made on the basis of VLBI observations. In the absence 
of VLBI observations, the source spectrum can be used as a proxy. 
\Note{It was found a long time ago \citep{r:ken69} that sources with 
a spectral index flatter than $-0.5$ tend to be compact as they are 
dominated by emission from the core}. The problem is that existing 
catalogues are not complete enough to determine the spectrum in the range 
of flux densities from 1--100~mJy, especially in the southern hemisphere.

  These considerations prompted us to propose a program of observations
of {\it all} unassociated 2FGL sources. The eventual goal of the program
is to find all AGNs with correlated flux densities brighter than 2~mJy 
at 8~GHz from regions smaller than 50~mas within the areas of 99 per cent 
probability of \Fermi\ localisation uncertainties. Association with 
a VLBI source automatically improves positions of $\gamma$-ray objects 
from arcminutes to milliarcseconds, allowing for an unambiguous 
association with sources at other wavelengths, for instance, optical. 
Combined with results of on-going programs to find pulsars associated 
with \Fermi\ sources, for instance \citet{r:barr}, we expect the 
list of unassociated sources to shrink. At the moment we do not know 
whether the list will shrink to zero or a population of radio-quiet 
$\gamma$-ray sources will be found.

  In this paper we report early results of the first step of the program 
of observations of \Fermi\ unassociated sources: observations with 
the Australia Telescope Compact Array (ATCA) of 411 sources with 
$\delta < +10\degr$. A brief overview of the program is given in 
section~\ref{s:observations}. The data analysis procedure is described
in section~\ref{s:analysis}. Results, the catalogue of 375 potential 
associations for 2FGL sources is given in section ~\ref{s:results}
followed by concluding remarks.

\section{Observations}
\label{s:observations}

\subsection{Observing Program}

\begin{figure}
   \includegraphics[width=0.48\textwidth]{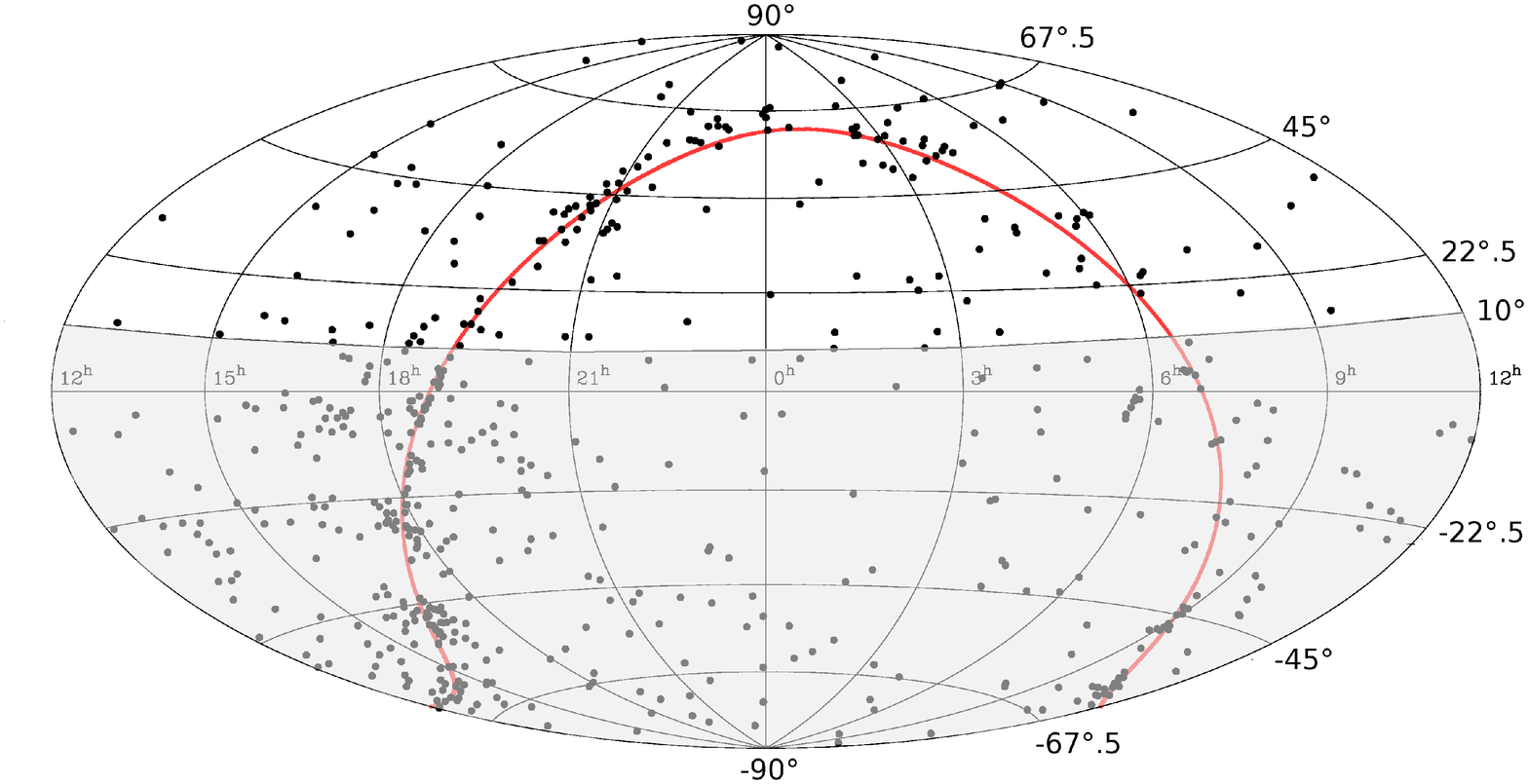}
   \caption{The distribution of 573 2FGL unassociated sources.
            The area of this survey is shown with grey colour.}
   \label{f:aofus_map}
\end{figure}

  The areas of 99 per cent probability of 2FGL localisation,
which corresponds to $3.035\sigma$, are typically 4--6~arcmin.
They are still too large for a blind search with VLBI using 
traditional approaches. Therefore, we organised observations 
in several stages. In the first stage we observed the sources 
with connected interferometers, the Very Large Array (VLA) in the 
northern hemisphere, and ATCA in the southern hemisphere. 
Observations are made in the remote wings of the wide-band receiving
system that covers 4.5--10.0~GHz. The sky distribution of 
target sources is presented in Figure~\ref{f:aofus_map}.

   We split the source list into two parts. A list of 411 sources with
declinations lower than $+10\degr$ was observed with the ATCA. 
A list of 215 sources with declinations higher than $0\degr$ was observed 
with the VLA. This list includes 35 overlapping sources in the declination 
band $[0, +10\degr]$ and 18 sources tentatively associated with supernova 
remnants and pulsar wind nebulae. Results of the VLA observations will be 
reported in a separate paper (F.~Schinzel et al., (2013) in preparation).

  Observations at two frequencies allow us to determine spectral indices. 
The spectral index is defined as $S \sim f^{\alpha}$, where $S$ is the flux 
density, and $f$ is the frequency. Sources with a spectral index $\alpha$ 
greater than $-0.5$ are classified as flat spectrum sources and are therefore
considered as likely associations. Bright detected sources will be followed 
up with the VLBA and LBA. Detection of emission at parsec scales will 
firmly associate them with the compact core regions of AGNs.

\subsection{ATCA observations}

  We observed a list of 411 target sources with ATCA on September 19--20
2012 for 29 hours. We used the automatic scheduling software {\sf sur\_sked}
originally developed for VLBI survey experiments~\citep{r:lcs1}. In order 
to maximise the observing efficiency, we reduced the correlator cycle time
from the usual 10 seconds to 6 seconds, and observed in the mosaic mode 
to minimise correlator overheads when changing sources. We tried to
observe each target source in three scans of 24 seconds each, separated 
by at least three hours in a sequence that minimises slewing time. In fact,
we observed 80 sources in 4 scans, 239 sources in 3 scans and 92 sources
in 2 scans. At the beginning of the experiment we observed the primary 
amplitude calibrator 1934$-$638 for 20 minutes. The automatic scheduling 
process checked that for every target source a phase calibrator was inserted 
into the schedule that satisfied two criteria: that the arc between the 
target and the calibrator should be less than $20\degr$ and a calibrator 
should be observed within 20 minutes of the target. We used a pool of 1464 
compact sources with $\delta < +20\degr$ and correlated flux densities 
$>200$~mJy. The scheduling process picked up 207 sources from this pool. 
Many adjacent target scans reused the same phase calibrator. Each calibrator 
scan was observed for 18 seconds.
 
  The array was in the H214 hybrid configuration with baselines 
ranging from 31--214 meters between the inner 5 antennas and $\sim$4.4~km 
between CA06 and the inner antennas. We observed simultaneously in two bands, 
both 2GHz wide, centred on 5.5 and 9.0 GHz. During data reduction the band 
edges were excluded and the ranges [4.58, 6.42] GHz (hereafter 5 GHz band) 
and [8.09, 9.92] GHz (hereafter 9 GHz band) were used. The data were recorded 
in both polarizations with the Compact Array Broadband Backend \citep{r:cabb}.

\section{Data Analysis}
\label{s:analysis}

  The ATCA correlator provided us the output\footnote{Available at 
\web{http://astrogeo.org/v1/aofusrpfits}} with frequency resolution 1~MHz
and time resolution 6~s. We used the software package Miriad 
\citep{r:miriad} for development of the analysis pipeline. 
The data processing carried out the following steps:

\renewcommand{\theenumi}{(\alph{enumi})}
\begin{enumerate}
   \item We split the data into subbands and into scans. We discarded
         data from the remote \Note{CA06} station for imaging analysis.

   \item We analysed the observations of all the sources and flagged 
         the data affected by radio interference using the task 
         \task{pgflag}. 

   \item We used the Miriad task \task{mfcal} for the bandpass 
         solution, using the data of 1934$-$638 with a solution interval of
         2 minutes. This bandpass was applied to all the data.

   \item We determined the antenna gains and polarization leakage of phase
         calibrators using the Miriad task \task{gpcal} and we also 
         corrected the flux scale with the task \task{gpboot}. The
         complex gain factors were then applied to the target sources.

   \item All calibrated visibilities of a target source were merged 
         into one file. The visibilities were inverted using 
         the Miriad task \task{invert} using the multi-frequency synthesis 
         algorithm. The typical image size was $1024 \times 1024$ with 
         a pixel size of 2.4~arcsec although for some target sources 
         we changed the pixel size later during the data analysis.  
         We tapered the data according to Brigg's visibility 
         weighting robustness parameter factor 0.5 during inversion.

   \item We cleaned the data using the Miriad task \task{mfclean}.
         The default CLEAN gain was set to 0.01 and the default
         maximum number of iterations was set to 192. The default
         clean region was 93 per cent of the image.

   \item We searched for point sources in the cleaned image using the task 
         \task{imsad}, requiring the flux densities to be greater than 5.5 
         times the image noise root mean square (rms).

   \item \Note{We performed the non-linear least square fit for positions and 
         parameters of the Gaussian models of detected sources to calibrated 
         visibilities using the Miriad task \task{uvsfit}. This task also
         computed the spectral index. We used positions and flux 
         densities of all sources found by \task{imsad}, excluding those 
         flagged as spurious, as the initial values for the fit 
         to the visibility data. A detailed description of the procedure is 
         given below}. We ran this procedure for each subband separately.

   \item \Note{For sources detected in both subbands we computed spectral indices 
               between 5~GHz and 9~GHz.}

\end{enumerate}

\Note{
The task \task{uvsfit} estimates source parameters through the minimisation 
of the least-squared differences between the measured visibilities and
those computed from a model source or sources.
} 

\note{
Given a set of observed visibilities, each measured at a point $(u,v)$ 
that is the projected baseline vector expressed in wavelengths, 
and at a frequency $f$, we write $ V_i = V(u_i,v_i,f_i)$. There are 
$n = n_b \times n_f$ visibility measurements, where $n_b$ and $n_f$ are 
the number of baselines and the number of frequency channels respectively.

Given a model source, we can compute model visibilities
$ M_i = M(u_i,v_i,f_i) $ at each point in $(u,v,f)$ space for which 
we have visibility measurements.  The model visibilities can be 
calculated as the sum of visibilities for several model sources.
The fitting process is the adjustment of model source parameters 
to minimise the quantity $\sum | V_i - M_i |^2$.  Both tasks use 
the Levenberg-Marquardt algorithm for function minimisation.
The diagonal terms of the covariance matrix are returned as the 
variances in fitted parameter values. 


For a point source with flux-density $S$ and position $(l,m)$ relative 
to the phase centre, \task{uvfit} computes each model visibility as:
$M(u,v) = S (\cos{\theta} + i \sin{\theta})$, where 
$ \theta = 2\pi(u l + v m)$.

   In \task{uvsfit}, the flux-density is expressed as a function 
of frequency, and new parameters are introduced to describe the spectral 
shape.  The model source is now a function of $S_0$, the flux-density 
at a reference frequency $f_0$, position relative to
phase centre $(l,m)$, and spectral shape parameters 
$(\alpha_0,\alpha_1,\alpha_2)$.  The model visibilities
are then:
\beq
   \begin{array}{l @{\;} l}
      M(u,v,f) & =  S(f) (\cos \theta + i \sin \theta) \\
      \theta   & =  2\pi(u l + v m) \\
      S(f)     & =  S_0 \left( \Frac{f}{f_0}\right) ^\alpha \\
      \alpha   & =  \alpha_0 + \alpha_1 (f - f_0) + \alpha_2 (f - f_0)^2.
   \end{array}
\eeq{e:m1}

Model visibilities for the two-dimensional Gaussian source with 
flux-density $S(f)$, position $(l,m)$, major and minor axes and 
position angle $(a,b,\phi)$, and spectral shape described 
by $(\alpha_0,\alpha_1,\alpha_2)$ are calculated by \task{uvsfit} as:
\beq
   \begin{array}{l @{\;} l}
      M(u,v,f) & =  S(f) (\cos{\theta} + i \sin{\theta}) \exp( -\frac{\pi^2}{4 \log 2}
      \beta)\\
      \theta & =  2\pi(u l + v m) \\
      S(f) & = S_0 \left( \frac{f}{f_0}\right) ^\alpha \\
      \alpha & = \alpha_0 + \alpha_1 (f - f_0) + \alpha_2 (f - f_0)^2 \\
      \beta & =  (b(u \cos \phi - v \sin \phi))^2 +  (a(u \sin \phi + v \cos \phi))^2.
   \end{array} \hspace{-3em}
\eeq{e:m2}
} 

   In our analysis we estimated only parameter $\alpha_0$

\begin{figure}
   \includegraphics[width=0.48\textwidth]{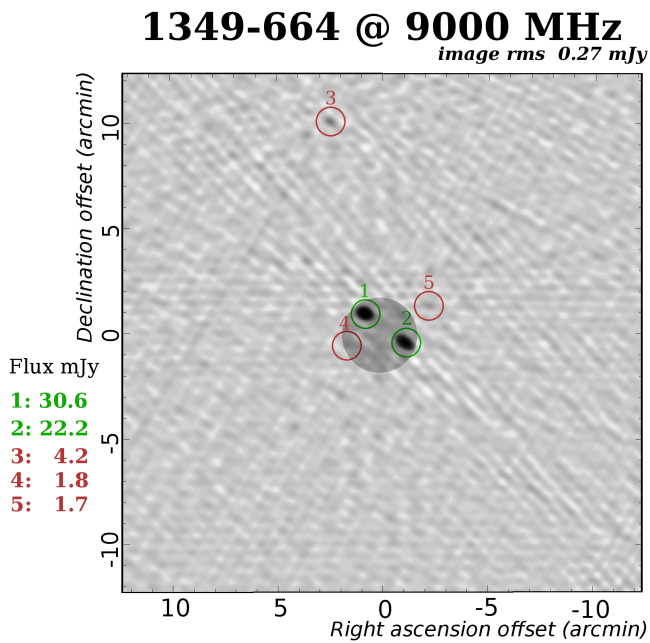}
   \caption{Example of a 9~GHz image of around 2FGL source J1353.5$-$6640.
            Flux densities were not corrected for beam attenuation.
            Objects 3--8 were flagged out. The gray area in the center
            of the image shows the error ellipse with the 
            probability 99 per cent to find there J1353.5$-$6640.
            }
   \label{f:1349-664}
\end{figure}

   At this point we visually checked every cleaned image. We have 
developed a web-application that allows us to interactively examine 
each image and flag those point sources found with 
{\sf imsad} that appear to be spurious or lie far away from 
the pointing direction. For 7~per cent of the images we manually 
adjusted parameters for inversion and cleaning, such as image 
size, pixel size, offset of image center, and the number of 
CLEAN iterations. Figure~\ref{f:1349-664} provides 
an example of one of the ATCA images\footnote{All 822 images derived 
from analysis of these observations are available online at 
\web{http://astrogeo.org/aofus}}. Sources 4--8 have 
a signal-to-noise ratio (SNR) in the range 5.5--6.7 --- not 
sufficient for a reliable detection in the absence of other 
information. There are two sources within the 99~per cent 
localisation error ellipse of J1353.5$-$6640. Source~1 has a 
spectral index of $-$0.33, while source~2 has a spectral index of $-$1.00. 
\Note{Source 1 was also detected in the ATPMN survey \citep{r:atpmn}
as J135340.1$-$663957 with flux densities $43 \pm 6$ and 
$28 \pm 10$ mJy at 4.8 and 8.6~GHz respectively.}
There is also a strong X-ray source 1RXS J135341.1$-$664002 within 
7~arcsec of source~1. \citet{r:tsa} observed its optical counterpart 
with the VLT,, which is within 1.7~arcsec of source~1, and reported 
a featureless spectrum. This suggests its classification as 
a BL~Lac type. This additional information presents a strong 
argument in favour of association of source~1 with 
2FGL J1353.5$-$6640.

  Source~3 was flagged out because it lies 10.3~arcmin from 
the pointing direction. The beam power at 9~GHz drops by 
a factor of 2 at 2.7~arcmin and has the first minimum at 6.5~arcmin. 
In fact, source 3 lies within 1~arcsec, i.e. within its position 
error, from J1353-6630 detected in the Australia Telescope 20 GHz Survey 
(AT20G) carried out with ATCA at 20~GHz~\citep{r:at20g}. The AT20G 
catalogue reports a flux density of 322 mJy at 8~GHz. The flux density 
measured from our observations is 77~times weaker because it was detected 
by a sidelobe of the beam. We did not try to calibrate source flux 
densities for objects that had offsets from the pointing 
direction beyond where the beam power falls below 0.2;
400~arcsec at 5~GHz and 250~arcsec at 9~GHz. For sources that 
lay within these limits we corrected their flux densities for 
beam attenuation $P(\theta)$ using an approximation of the beam 
pattern measured by \citet{r:atca_beam} in the form of
\beq
  P(\theta) = \Frac{1}{\dss\sum_{k=0}^{k=4} C_k (\theta \, f)^{2k}}
\eeq{e:e5}
   where $\theta$ is the distance from the pointing direction,
$f$ is the observing frequency, and $C_k$ are coefficients that 
approximate beam pattern measurements.

\section{Results}
\label{s:results}

\begin{table*}
   \caption{The first 8 rows of 146 objects detected at both 5 
            and 9.0~GHz. The table columns are explained 
            in the text. The full table is available in the 
            electronic attachment.}
   \tiny
   \begin{tabular}{ l @{\hhem} l @{\hhem} l @{\hhem} l @{\hhem} r 
                      @{\hhem}  r @{\hhem} r @{\hem}  r @{\hhem} r 
                      @{\hhem} r @{\hhem} r @{\hem}  r @{\hem}  r 
                      @{\hhem} r @{\hhem} r @{\hem}  r @{\hem}  r 
                      @{\hem}  r @{\hem}  r @{\hhem} r @{\hhem} c 
                      @{\hhem} r @{\hhem} r}
      \hline
         \fs 2FGL name &
         \fs IAU name  &
         \fs Fl        &
         \ntab{c}{\fs $\alpha$} &
         \ntab{c}{\fs $\delta$} &
         \fs $\Delta \alpha$    &
         \fs $\Delta \delta$    &
         \fs $F_5$ &
         \fs $F_9$ &
         \fs $\sigma_5$ &
         \fs $\sigma_9$ &
         \fs Sp${}_5$ &
         \fs Sp${}_9$ &
         \fs $\sigma$Sp${}_5$ &
         \fs $\sigma$Sp${}_9$ &
         \fs Sp &
         \fs $\sigma$Sp &
         \fs D &
         \fs N$\sigma$ &
         \fs $F_1$ &
         \fs $C_1$ &
         \ntab{c}{\fs 1GHz ID} &
         \ntab{c}{\fs WISE ID} \\
         & 
         &
         &
         hr ~min ~sec & 
         ${}^\circ \hspace{1.0em} ' \hspace{1.0em}  ''$ &
         \ntab{c}{$''$} &
         \ntab{c}{$''$} &
         mJy &
         mJy &
         mJy &
         mJy &
             &
             &
             &
             &
             &
             &
         $'$ &
         $'$ &
         mJy &
             &
             &
         \vspace{0.5ex} \\
         \ntab{c}{(1)}    &
         \ntab{c}{(2)}    &
         (3)              &
         \ntab{c}{(4)}    &
         \ntab{c}{(5)}    &
         (6)    &
         (7)    &
         (8)    &
         (9)    &
         (10)    &
         (11)    &
         (12)    &
         (13)    &
         (14)    &
         (15)    &
         (16)    &
         (17)    &
         (18)    &
         (19)    &
         (20)    &
         (21)    &
         \ntab{c}{(22)}  &
         \ntab{c}{(23)}    
         \vspace{0.5ex} \\
      \hline
      J0031.0$+$0724 & J0031+0724 &  f & 00 31 19.69 & $+$07 24 53.8 & 0.8 & 1.0 &    9.6 &    7.5 &  0.4 &  1.0 & $-$0.21 & $-$9.90 & 0.24 & $-$9.90 & $-$0.49 & 0.29 &  3.4 & 1.5 &   11.6 & N & 003119$+$072456 & J003119.70$+$072453.6 \\
      J0102.2$+$0943 & J0102+0944 &a f & 01 02 17.12 & $+$09 44 10.4 & 0.6 & 0.8 &   14.7 &   16.4 &  0.3 &  0.4 & $ $0.40 & $-$0.11 & 0.12 & $ $0.26 & $ $0.22 & 0.07 &  1.2 & 0.4 &   11.0 & N & 010217$+$094407 &                       \\
      J0116.6$-$6153 & J0116-6153 &a f & 01 16 19.70 & $-$61 53 43.0 & 0.5 & 0.8 &   33.5 &   30.5 &  0.3 &  0.6 & $-$0.04 & $-$0.30 & 0.05 & $ $0.20 & $-$0.19 & 0.05 &  2.7 & 1.4 &   24.0 & S & 011619$-$615343 & J011619.59$-$615343.5 \\
      J0116.6$-$6153 & J0116-6156 &  f & 01 16 43.97 & $-$61 56 53.3 & 0.6 & 0.8 &   17.5 &   15.8 &  0.4 &  1.3 & $-$0.31 & $-$9.90 & 0.14 & $-$9.90 & $-$0.20 & 0.18 &  3.7 & 1.7 &   33.0 & S & 011643$-$615653 &                       \\
      J0116.6$-$6153 & J0116-6150 &    & 01 16 56.83 & $-$61 50 12.5 & 0.9 & 1.1 &    7.1 &    5.4 &  0.4 &  1.2 & $-$1.04 & $-$9.90 & 0.29 & $-$9.90 & $-$0.54 & 0.47 &  3.5 & 1.6 &   43.0 & S & 011656$-$615013 &                       \\
      J0143.6$-$5844 & J0143-5845 &a f & 01 43 47.45 & $-$58 45 51.8 & 0.5 & 0.8 &   24.0 &   22.3 &  0.2 &  0.4 & $-$0.18 & $-$0.47 & 0.06 & $ $0.17 & $-$0.15 & 0.04 &  1.5 & 1.2 &   26.0 & S & 014347$-$584550 & J014347.39$-$584551.3 \\
      J0316.1$-$6434 & J0316-6437 &  f & 03 16 14.11 & $-$64 37 30.8 & 0.6 & 0.9 &   15.7 &   12.5 &  0.4 &  1.0 & $ $0.18 & $-$9.90 & 0.15 & $-$9.90 & $-$0.46 & 0.16 &  3.5 & 1.5 &   12.0 & S & 031614$-$643732 & J031614.31$-$643731.4 \\
      J0409.8$-$0357 & J0409-0400 &a f & 04 09 46.58 & $-$04 00 02.4 & 0.5 & 0.8 &   89.9 &   87.7 &  0.3 &  0.7 & $ $0.35 & $-$0.56 & 0.02 & $ $0.08 & $-$0.05 & 0.02 &  2.4 & 1.1 &   39.1 & N & 040946$-$040003 & J040946.57$-$040003.4 \\
      \hline
   \end{tabular}
   \label{t:table1}
\end{table*}

  In summary, we detected 571 objects in 338 fields. Of these, 
146 objects have position offsets beyond the areas 
of 99 per cent probability of the 2FGL positions. These objects were 
discarded from further analysis.

  The rms of the images is typically in the range of 0.15--0.25~mJy.
The FWHM size of the restored beam is typically 35 arcsec at 5~GHz 
and 20~arcsec at 9 GHz. The detection limit is 1.8~mJy for sources 
in the center of the field of view and 9~mJy at the edge of the field 
of view, at 6.5~arcmin. There are 23 fields that have a number of resolved 
objects, for instance the field around 2FGL source J1619.7$-$5040 which 
corresponds to the HII region G332.8$-$0.6 \citep{r:hii}. Although the 
data analysis pipeline selected sources that it considers as ``points'', 
these are actually hot spots in an extended Galactic object rather than 
separate compact extragalactic sources. Imaging of an extended object 
using 2--4 scans with a five-element array gives inconclusive results. 
We flagged sources from such fields as possibly extended. Extra care 
must be taken when dealing with these objects.

  We cross-matched all remaining sources against the NVSS \citep{r:nvss}, 
SUMSS version 2.1 of 2012 February 16 \citep{r:summs1,r:summs2}, 
MGPS-2 \citep{r:mgps2}, and WISE \citep{r:wise} catalogues. The NVSS 
catalogue is derived  from VLA observations at 1.4~GHz and the SUMSS and 
MGPS-2 catalogues are derived from observations at 0.843~GHz with 
the Molonglo Observatory Synthesis Telescope. These catalogues have 
similar resolutions of $\sim\!40$~arcsec and are complimentary to each 
other as they cover different areas of the sky. We used a search radius of 
20~arcsec to find counterparts in the NVSS, SUMSS and MGPS-2 catalogues, 
and $2\sigma$ position uncertainties of the ATCA coordinates for 
matching to a WISE object.

  We have 160 matches in the WISE catalogue, 193 matches in NVSS, 
59 matches in SUMSS, and 54 matches in MGPS-2.
Since the average positional error of WISE is rather small, 
0.25~arcsec, the mathematical expectation of the number of background 
WISE sources that fall within the $2\sigma$ error ellipse of ATCA position 
(in the case that the ATCA detections and WISE objects are physically 
unrelated) is only 10. We found matches for 16 times 
more sources, roughly 1/3 of our list. Thus, we conclude that 
the majority of the ATCA--WISE matches are real.

  The NVSS catalogue contains objects with declinations $>-40\degr$.
We have 243 ATCA detections with $\delta > -40\degr$ and 191 of these, 
i.e. 77 per cent, have a counterpart in NVSS. Among the 181 ATCA detected
objects with declinations $<-40\degr$, 63 per cent have counterparts 
in SUMSS and MGPS-2. We have less counterparts with declinations below 
$-40\degr$ because these catalogues are not as deep as NVSS: their 
limiting peak brightness is 6~mJy beam${}^{-1}$ at $\delta \le -50\degr$ 
and 10~mJy beam${}^{-1}$ at $\delta > -50\degr$, while NVSS has 
a limiting peak brightness 2.5~mJy.

  We used these matches to re-calibrate position errors. We selected
92 matches that have flux densities brighter than 15~mJy at 1.4~GHz. NVSS 
positions of such sources are accurate to within less than 1~arcsec.
We formed the dataset of position differences normalised to their
standard deviations, which is the sum in quadrature of NVSS 
position uncertainties and position uncertainties from our ATCA 
observations. We used a reweighting model of the form 
$\sqrt{ (r\,\sigma)^2 + f^2}$ to bring the averaged normalised 
residuals close to one. We found the following
scale-factors and error floors: 1.45 and 0.5~arcsec for right ascension
and 1.80 and 0.75~arcsec for declination. We inflated our position errors
according to this reweighting model.

\begin{figure}
   \includegraphics[width=0.48\textwidth]{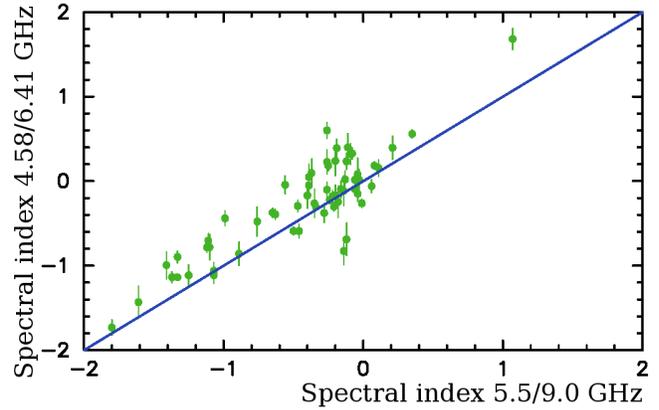}
   \caption{Comparison of source spectral indices from dual sub-band
            observations (horizontal axis) with the spectra
            determine from the low sub-band only (vertical axis).
            }
   \label{f:spind}
\end{figure}

\Note{
  For all detected sources we determined the spectral index within
the subband. In addition, for sources detected at both subbands independently, 
we determined spectral indices between 5.5 and 9 GHz from their flux densities. 
In general, the spectral index between both subbands and from the low subband 
(5.5 GHz) spectra do not necessarily coincide because systematic errors
affect the estimates in different ways and some sources may have
spectra that deviate from the power law. In order to compare the consistency
of spectral index estimates, we computed spectral indices using the low 
subband spectra for the sources that were detected in both bands as well.
The scatter plot of this comparison is shown in Figure~\ref{f:spind}.
\Note{We see that the spectral index across the [4.58, 6.42] GHz subband 
is slightly higher than the spectral index across [4.58, 9.91] GHz. This can 
be explained by a curvature in the spectrum.} Results of this comparison 
demonstrate a reasonable consistency of spectral indices estimates using 
both approaches.
}   

  In order to evaluate the significance of these $\gamma$-ray associations, 
we computed the likelihood ratios $\Lambda$. This is defined as the ratio 
of the probability that a radio counterpart will be found {\it inside} a disk 
of radius $d$ to the probability to find a background radio source with 
flux $S$ or greater {\it outside} the same disk:
\beq
      \Lambda = \Frac{e^{-n^2/2}}{N(1) \, S^{\,p} d^2/4}
\eeq{e:6}
  where $n$ is the normalised arc $d$ between the radio and the 2FGL position,
$N(1) and p$ are parameters of the $\log N(\log S)$ dependence that
describes the number of sources $N$ with radio flux densities 
greater than $S$. 

Among large radio surveys, the 4.85~GHz GB6 \citep{r:gb6}
catalogue, with a FWHM of the beam of 3~arcmin, is the 
closest match to our ATCA observations at 5~GHz. We built a cumulative 
histogram log N--log S over 75,162 GB6 sources at 6.06~sr and fit 
a straight line to it in the interval of flux densities 
[30, 810]~mJy. We obtain $N(1 {\rm Jy})=965$ and $p=-1.3475$ from this 
analysis. We should note that these parameters are different than
those we derived from the analysis of VLBI results because they are
related to a different population of radio sources. The likelihood 
ratio describes our knowledge of how likely a radio source with 
a given flux density can be found at a given distance by chance. 
Sources with large $\Lambda$ are less likely located close to 
the $\gamma$-ray counterpart by chance.

  The list of 424 sources falls into three categories:

\begin{table*}
   \caption{The first 8 rows of 229 objects detected in the 5~GHz subband only.
            The table columns are explained in the text. The full table
            is available in the electronic attachment.}
   \tiny
   \begin{tabular}{ l @{\hem}  l @{\hhem} c @{\hhem} l @{\hhem} r @{\hem} r
                      @{\hem}  r @{\hem} r @{\hhem}  r @{\hem}  r 
                      @{\hem}  r @{\hem} r @{\hem}  r @{\hem} r @{\hhem}  c @{\hhem} r @{\hem} r}
      \hline
         \fs 2FGL name &
         \fs IAU name  &
         \fs Fl        &
         \ntab{c}{\fs $\alpha$} &
         \ntab{c}{\fs $\delta$} &
         \fs $\sigma \alpha$    &
         \fs $\sigma \delta$    &
         \fs $F_5$ &
         \fs $\sigma_5$ &
         \fs Sp &
         \fs $\sigma$Sp &
         \fs D &
         \fs N$\sigma$ &
         \fs $F_1$ &
         \fs $C_1$ &
         \ntab{c}{\fs 1GHz ID} &
         \ntab{c}{\fs WISE ID} \\
         & 
         &
         &
         hr ~min ~sec & 
         ${}^\circ \hspace{1.0em} ' \hspace{1.0em}  ''$ &
         \ntab{c}{$''$} &
         \ntab{c}{$''$} &
         mJy &
         mJy &
             &
             &
         $'$ &
         $'$ &
         mJy &
             &
             &
         \vspace{0.5ex} \\
         \ntab{c}{(1)}    &
         \ntab{c}{(2)}    &
         \ntab{c}{(3)}    &
         \ntab{c}{(4)}    &
         \ntab{c}{(5)}    &
         (6)              &
         (7)              &
         (8)              &
         (9)              &
         (10)             &
         (11)             &
         (12)             &
         (13)             &
         (14)             &
         (15)             &
         \ntab{c}{(16)}   &
         \ntab{c}{(17)}    
         \vspace{0.5ex} \\
      \hline
      J0014.3$-$0509 & J0014$-$0512 &   & 00 14 33.84 & $-$05 12 48.0 & 3.3 & 3.8 &    2.7 & 0.6 & $-$9.90 & $-$9.90 &  5.1 & 1.5 &    4.3 & N & 001433$-$051244 & J001433.86$-$051249.6  \\
      J0158.4$+$0107 & J0158$+$0108 &   & 01 58 41.88 & $+$01 08 23.7 & 3.2 & 4.5 &    3.0 & 0.7 & $-$9.90 & $-$9.90 &  4.2 & 1.1 &   17.6 & N & 015842$+$010817 &                        \\
      J0200.4$-$4105 & J0200$-$4109 &   & 02 00 20.91 & $-$41 09 37.2 & 1.5 & 1.7 &    4.8 & 0.5 & $-$9.90 & $-$9.90 &  3.9 & 1.7 &        &   &                 & J020020.94$-$410935.6  \\
      J0226.1$+$0943 & J0226$+$0937 &   & 02 26 13.81 & $+$09 37 26.9 & 0.5 & 0.8 &  157.7 & 0.9 & $-$0.22 & $ $0.03 &  5.8 & 2.5 &  374.6 & N & 022613$+$093726 & J022613.70$+$093726.8  \\
      J0312.5$-$0914 & J0312$-$0914 &   & 03 12 16.22 & $-$09 14 17.5 & 3.1 & 2.4 &   11.3 & 1.5 & $-$0.13 & $ $0.29 &  4.4 & 1.5 &   39.2 & N & 031216$-$091421 & J031216.17$-$091418.7  \\
      J0312.5$-$0914 & J0312$-$091A &   & 03 12 18.26 & $-$09 14 35.7 & 5.1 & 3.7 &    5.1 & 1.5 & $-$9.90 & $-$9.90 &  3.9 & 1.4 &        &   &                 & J031218.15$-$091436.4  \\
      J0312.5$-$0914 & J0312$-$0919 &   & 03 12 21.67 & $-$09 19 01.4 & 2.0 & 2.2 &    5.8 & 0.7 & $-$9.90 & $-$9.90 &  5.0 & 2.1 &   10.6 & N & 031221$-$091906 & J031221.58$-$091902.8  \\
      J1046.8$-$6005 & J1046$-$6005 &ae & 10 46 17.27 & $-$60 05 21.6 & 0.5 & 0.8 &  264.2 & 1.8 & $-$0.92 & $ $0.03 &  4.0 & 1.6 &        &   &                 &                        \\
      \hline
   \end{tabular}
   \label{t:table2}
\end{table*}

\begin{table*}
   \caption{The first 8 rows of 49 objects detected beyond 6.5 arcmin of 
            the 5~GHz pointing centre, or detected only at 9~GHz. The table columns 
            are explained in the text. The full table is available 
            in the electronic attachment.}
   \begin{tabular}{ l @{\hem}  l @{\hhem}  c @{\hhem} l @{\hhem} 
                    r @{\hem}  r @{\hem}   r @{\hem}  r @{\hem}  
                    r @{\hem}  r @{\hhem}  c @{\hhem} r @{\hem}  
                    r @{\hem} }
      \hline
         \ns 2FGL name          &
         \ns IAU name           &
         \ns Fl                 &
         \ntab{c}{\ns $\alpha$} &
         \ntab{c}{\ns $\delta$} &
         \ns $\sigma \alpha$    &
         \ns $\sigma \delta$    &
         \ns D                  &
         \ns N$\sigma$          &
         \ns $F_1$          &
         \ns $C_1$          &
         \ntab{c}{\ns 1GHz ID}  &
         \ntab{c}{\ns WISE ID}  \\
         & 
         &
         &
         hr ~min ~sec & 
         ${}^\circ \hspace{1.0em} ' \hspace{1.0em}  ''$ &
         \ntab{c}{$''$} &
         \ntab{c}{$''$} &
         $'$            & 
         $'$            &
         mJy            &  
                        &  
                        &  
         \vspace{0.5ex} \\
         \ntab{c}{(1)}  &
         \ntab{c}{(2)}  &
         \ntab{c}{(3)}  &
         \ntab{c}{(4)}  &
         \ntab{c}{(5)}  &
         (6)            &
         (7)            &
         (8)            &
         (9)            &
         (10)           &
         (11)           &
         \ntab{c}{(12)} &
         \ntab{c}{(13)}    
         \vspace{0.5ex} \\
      \hline
         J0414.9$-$0855 & J0415$-$0854 &   & 04 15 23.27 & $-$08 54 26.5 & 3.6 & 3.5 & 6.9 & 2.1 &  10.3 & N & 041523$-$085425 & J041523.13$-$085424.4 \\
         J0540.1$-$7554 & J0539$-$7601 &   & 05 39 18.29 & $-$76 01 30.6 &13.7 & 4.4 & 8.0 & 2.2 &  79.0 & S & 053916$-$760131 & J053916.48$-$760130.3 \\
         J0540.1$-$7554 & J0540$-$7601 &   & 05 40 10.51 & $-$76 01 54.0 &11.8 & 3.9 & 7.7 & 2.2 &  37.0 & S & 054011$-$760156 & J054011.59$-$760154.3 \\
         J0540.1$-$7554 & J0541$-$7601 &   & 05 41 22.48 & $-$76 01 09.6 &18.9 & 5.8 & 8.1 & 2.4 &  31.0 & S & 054123$-$760108 & J054123.16$-$760107.6 \\
         J0547.5$-$0141 & J0547$-$0133 &   & 05 47 20.86 & $-$01 33 30.9 & 1.4 & 1.8 & 8.1 & 1.4 &  22.1 & N & 054720$-$013327 & J054720.85$-$013329.9 \\
         J0555.9$-$4348 & J0555$-$4346 &   & 05 55 16.04 & $-$43 46 30.7 & 1.8 & 1.4 & 7.1 & 2.2 & 144.0 & S & 055516$-$434631 &                       \\
         J0555.9$-$4348 & J0555$-$4345 &   & 05 55 19.57 & $-$43 45 28.9 & 2.2 & 1.7 & 6.8 & 2.1 & 101.0 & S & 055519$-$434534 &                       \\
         J0600.8$-$1949 & J0600$-$1950 & a & 06 00 59.64 & $-$19 50 39.4 & 1.0 & 1.0 & 2.0 & 0.5 &  96.7 & N & 060100$-$195049 &                       \\
      \hline
   \end{tabular}
   \label{t:table3}
\end{table*}

\renewcommand{\theenumi}{\bf(\arabic{enumi})}
\begin{enumerate}
  \item {\bf Category I:} 146 objects detected in both the 5~GHz and 
        9.0~GHz subbands within 2.7~arcmin of the pointing direction. 
        We provide in Table~\ref{t:table1} the $\gamma$-ray source 
        name; IAU name of the detected radio source; tentative association 
        status, estimates of its J2000 coordinates followed by 
        $1\sigma$ uncertainties ($\sigma \alpha$ and $\sigma \delta$) 
        in arcseconds (the uncertainties in Right 
        Ascension are not scaled by $\cos\delta$); flux densities at 5~GHz 
        and 9~GHz in mJy ($F_5$ and $F_9$) corrected for beam 
        attenuation followed by their standard deviations 
        ($\sigma_5$ and $\sigma_9$); spectral indices within the 5 and 9
        GHz subbands Sp${}_5$, Sp${}_9$, and their standard deviations,
        followed by the spectral index from 5--9 GHz (Sp and $\sigma$Sp) 
        computed from flux densities $F_5$ and $F_9$. Spectral index 
        estimates with uncertainties greater than 0.4 are omitted.
        We provide the distance of a source from the pointing direction $D$, 
        followed by $N\sigma$, the ratio of this distance to its standard 
        deviation derived from the reported 2FGL position localisation errors. 
        If the source was associated with an object either from NVSS, SUMSS,
        or MGPS-2 catalogues, its flux density at 1.4~GHz (NVSS) or 0.843~GHz
        (SUMSS and MGPS-2) is shown in column $F_1$ followed by the 1~GHz 
        catalogue code ($N$ for NVSS, $S$ for SUMSS, $M$ for MGPS-2), 
        and the source identifier in that catalogue. If the source 
        was associated with a WISE object, its WISE source ID is shown 
        in the last column.

        Column~3 shows two flags: ``a'' if the source has likelihood 
        ratio greater than 10, and therefore, considered a likely 
        association, ``e'' if the source is extended, and ``f'' if the source 
        has a spectral index flatter than $-0.5$.

  \item {\bf Category II:} 229 objects detected only at 5~GHz  
        within 6.5~arcmin of the pointing direction. We provide in 
        Table~\ref{t:table2} estimates of the flux density at 5~GHz corrected 
        for beam attenuation. The contents of Table~\ref{t:table2} is 
        similar to Table~\ref{t:table1}, except columns $F_9$, 
        $\sigma_9$, Sp, and $\sigma$Sp are excluded. The spectral 
        index is computed only over the sub-band [4.58, 6.41] GHz. 
        Spectral index estimates with uncertainties greater than 0.4 
        are omitted.
        
  \item {\bf Category III:} 49 objects either detected beyond 6.5~arcmin 
        of the pointing direction or detected only at 9~GHz. Since 
        calibration for beam attenuation becomes uncertain at large 
        distances, we can only provide a lower limit estimate of their flux 
        density: 20~mJy. Table~\ref{t:table3} lists these sources. 
        Its contents are similar to Table~\ref{t:table2}, except columns 
        $F_5$ and $\sigma_5$, spectral index and its uncertainty 
        are excluded.

        A fill value of $-9.9$ in all tables indicates a lack of information.
\end{enumerate}

\begin{figure}
   \includegraphics[width=0.48\textwidth]{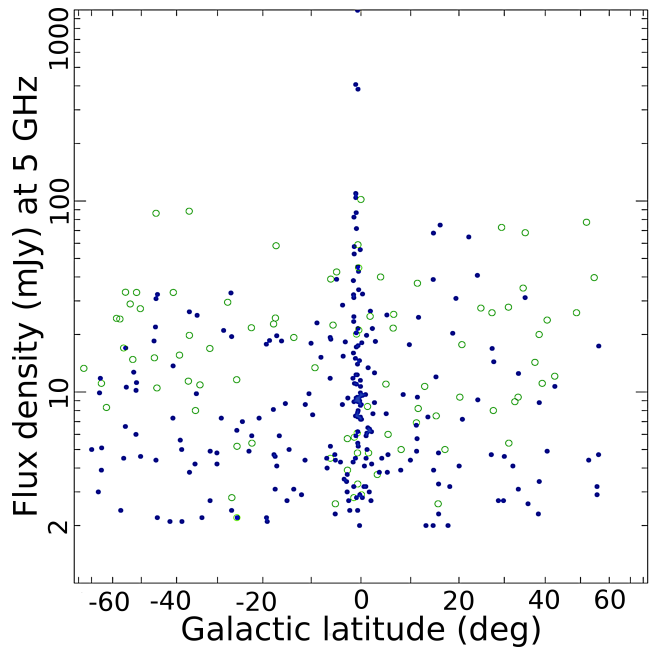}
   \caption{Dependence of the flux density of detected sources in
            categories I and II on galactic latitude. Flat
            spectrum sources are shown with green hollow circles. 
            Other sources are shown with filled blue discs.
            }
   \label{f:flux}
\end{figure}

As can be seen from Figure~\ref{f:flux}, many of the unassociated
\Fermi\ sources are at low Galactic latitude. For the 30\% of the
sources in Tables~\ref{t:table1}--\ref{t:table3} which lie within 
1~degree of the Galactic plane, contamination of our ATCA images 
by Galactic HII regions, supernova remnants and planetary nebulae 
is common so careful work will be needed in the final identification 
process. As expected, flat-spectrum sources do not exhibit
concentration to the Galactic plane.

\section{Concluding remarks}
\label{s:conclusions}

  We found 424 counterparts within 99 per cent probability of 
localisation of 267 2FGL sources (Tables~\ref{t:table1}--\ref{t:table3}). 
For 141 2FGL sources more than one counterpart was found. Among 424
counterparts, 84 have a spectrum flatter than $-0.5$, 62 have a spectrum 
steeper than $-0.5$, and for 278 objects no spectral index has been 
determined. The flat spectrum sources are considered as tentatively 
associated with \Fermi\ objects. For 128 sources the probability of 
association exceeds more than 10 times the probability to find 
a background source with that flux density by chance. These flags help 
to identify a subset of probable associations. However, taken alone,
they do not establish a firm association. For instance, object J1045$-$5941 
has a spectral index of 1.21, but it is a Galactic object ($\eta$ Carinae),
not a blazar. Some compact HII regions and planetary nebulae may 
have flat spectrum. Additional information, e.g., from observations 
at other wavelengths, will help to identify a source associated 
with a $\gamma$-ray object.

  Similarly, a steep spectrum does not rule out an extragalactic nature of 
an object. \citet{r:Fermi_VCS} presented several examples of 
steep-spectrum sources associated with \Fermi\ objects. Detecting 
emission on parsec-scales will allow us to determine the nature of
a radio source.

  There are 46 objects from the fields that have a number of 
resolved objects. Many of these fields are associated with HII
regions. Objects from these fields may in fact be hot spots of extended
objects, not separate sources. Only 4 of them, i.e. 9 per cent,
are associated with WISE objects, while 35 per cent of the remaining
sources have counterparts from the WISE catalogue.

  We did not find radio counterparts for 128 of our targets, roughly 1/3
of our list. Since for many sources the $3.035\sigma$ 2FGL position error
ellipse exceeded the field of view of 6.5~arcmin at 5~GHz, we cannot 
claim there is no source brighter than 10~mJy in the field.

  In the second phase of this project we plan to re-observe these 
fields in the mosaic mode to cover a larger area. This will allow us to 
detect all the sources in the 99 per cent probability of 2FGL localisation 
that are brighter than 2~mJy. We also plan to re-observe with ATCA those 
sources that were detected at 5~GHz, but are not detected at 9~GHz, and 
had position offsets exceeding the FWHM of the 9~GHz beam. To obtain more 
accurate positions and flux densities, we will also re-observe sources listed 
in Table~\ref{t:table3}, pointing to their positions found from
this survey. \Note{New observations are scheduled in September 2013.}

  In the third phase of the project we will observe detected sources
with the LBA to determine correlated flux densities from regions smaller
than 50~mas. This will allow us to associate detected radio sources 
with blazars.

\section{Acknowledgments}
\label{s:acknowledgments}

We wish to thank Mark Wieringa, Robin Wark, and Jamie Stevens 
for assistance in planning and carrying out the observations and processing 
the resulting data, and all the staff at the Paul Wild Observatory for 
their upkeep of the facility. We would like to thank James Condon and 
Greg Taylor for fruitful suggestions. F.K.S. acknowledges support by
the NASA Fermi Guest Investigator program, grant NNX12A075G.
The Australia Telescope Compact Array is part of the Australia Telescope 
National Facility which is funded by the Commonwealth of Australia for 
operation as a National Facility managed by CSIRO. This publication 
makes use of data products from the Wide-field Infrared Survey Explorer, 
which is a joint project of the University of California, and 
the JPL/California Institute of Technology, funded by the NASA.


\section{Supporing information}

Additional Supporting information may be found in the online version of this article:

{\bf Table S1.} The 146 objects detected at both 5 and 9.0 GHz. The
table columns are explained in the text.

{\bf Table S2.} The 229 objects detected in the 5-GHz sub-band only.
The table columns are explained in the text.

{\bf Table S3.} The 49 objects detected beyond 6.5 arcmin from the 5-
GHz pointing centre or detected only at 9 GHz. The table columns
are explained in the 
text\footnote{\web{http://mnras.oxfordjournals.org/lookup/suppl/doi:10.1093/mnras/stt550/-/DC1}}.

\label{lastpage}


\begin{thebibliography}{99}

   \bibitem[Ackermann et~al.(2012)]{r:fermi_agn} 
      Ackermann~M., et al., 2012, ApJ, 753, 83

   \bibitem[Barr et al.(2013)]{r:barr} 
      Barr~E.~D., et al.\
         2013, \mnras, 429, 1633

   \bibitem[Beasley et~al.(2002)]{r:vcs1}
      Beasley~A.~J., Gordon~D., Peck~A.~B., Petrov~L.,
      MacMillan~D.~S., Fomalont~E.~B., Ma~C.\
       2002, \apjs, 141, 13

   \bibitem[Bock, Large \& Sadler(1999)]{r:summs1}
     Bock~D.~C.-J., Large~M.~I., Sadler~E.~M., 
	1999, \aj, 117, 1578

   \bibitem[Condon et~al. (1998)]{r:nvss}
     Condon~J.~J., Cotton~W.~D., Greisen~E.~W., Yin~Q.~F.,
       Perley~R.~A., Taylor~G.~B., Broderick~J.~J.\
       1998, \aj, 115, 1693

   \bibitem[Condon et~al.(2011)]{r:v2m}
      Condon~J., Darling~J., Kovalev~Y.~Y., Petrov~L.\
        2011, arXiv:1110.6252

   \bibitem[Fomalont et~al.(2003)]{r:vcs2}
      Fomalont~E., Petrov~L., McMillan~D.S., Gordon~D., Ma~C.,
        2003, \aj, 126, 2562
       
   \bibitem[Gregory et~al.(1996)]{r:gb6}
     Gregory~P.~C., Scott~W.~K., Douglas~K., Condon~J.~J.\
        1996, \apjs, 103, 427

   \bibitem[Kellermann \& Pauliny-Toth(1969)]{r:ken69}
      Kellermann,~K.~I., \& Pauliny-Toth,~I.~I.~K.\ 
        1969, \apj, 155, L71 

   \bibitem[Kovalev et~al.(2007)]{r:vcs5}
     Kovalev~Y.Y., Petrov~L., Fomalont~E., Gordon~D.,
        2007, \aj, 133, 1236

   \bibitem[Kovalev et al.(2009)]{r:FM2}
      Kovalev~Y.~Y., et al.\ 
        2009, ApJ, 696, L17

   \bibitem[Kovalev(2009)]{r:Fermi_VCS}
     Kovalev~Y.~Y.\ 2009, ApJ, 707, L56

   \bibitem[Kuchar \& Clark(1997)]{r:hii}
     Kuchar~T.~A. \& Clark~F.~O. 1997, \apj, 488, 224

   \bibitem[Lister et al.(2009)]{r:Lister09}
     Lister~M.~L., Homan~D.~C., Kadler~M., Kellermann~K.~I., Kovalev~Y.Y., 
       Ros~E., Savolainen~T., Zensus~J. A.\
         2009, ApJ, 696, L22

   \bibitem[Mauch et al.(2003)]{r:summs2}
      Mauch~T., Murphy~T., Buttery~H.~J., Curran~J., 
      Hunstead~R.~W., Piestrzynska~B., Robertson~J.~G., 
      Sadler~E.~M. (2003), 
        \mnras, 342, 1117

   \bibitem[McConnell et al.(2012)]{r:atpmn}
      McConnell~D., Sadler~E.~M., Murphy~T., Ekers~R.~D. (2012),
        \mnras, 422, 1527

   \bibitem[Murphy et~al.(2007)]{r:mgps2}
     Murphy~T., Mauch~T., Green~A., Hunstead~R.~W., 
       Piestrzynska~B., Kels~A.~P., Sztajer~P.,
        2007, \mnras, 382, 382

   \bibitem[Murphy et~al.(2010)]{r:at20g} 
        Murphy~T. et~al. 2010, \mnras, 420, 2403

   \bibitem[Nolan et~al.(2012)]{r:2fgl} 
      Nolan~P.~L. et al., 2012, ApJS, 199, 31

  \bibitem[Petrov et~al.(2005)]{r:vcs3}
     Petrov~L., Kovalev~Y.Y., Fomalont~E., Gordon~D.,
       2005, \aj, 129, 1163

  \bibitem[Petrov et~al.(2006)]{r:vcs4}
     Petrov~L., Kovalev~Y.Y., Fomalont~E., Gordon~D.,
       2006, \aj, 131, 1872

   \bibitem[Petrov et al.(2008)]{r:vcs6}
    Petrov~L., Kovalev~Y.~Y., Fomalont~E.~B., Gordon~D.\
      2008, \aj, 136, 580 

  \bibitem[Petrov et~al.(2009)]{r:rdv}
     Petrov~P., Gordon~D., Gipson~J., MacMillan~D., Ma~C., Fomalont~E.,
       Walker~R.C., Carabajal~C.,
       2009, J~Geod, vol. 83(9), 859

  \bibitem[Petrov \& Taylor(2011)]{r:vips}
   Petrov, L., Taylor, G.~B.,\
      \aj, 2011, 142, 89

  \bibitem[Petrov et~al.(2011a)]{r:vgaps}
      Petrov~L., Kovalev~Y.~Y., Fomalont~E., Gordon~D.\
        2011a, \aj, 142, 35

  \bibitem[Petrov et al.(2011b)]{r:lcs1}
      Petrov~L., Phillips~C., Bertarini~A., Murphy~T., Sadler~E.~M.,
      2011b, \mnras, 414(3), 2528

   \bibitem[Petrov(2012)]{r:egaps}
      Petrov~L., 2012, \mnras, 416, 1097

   \bibitem[Pushkarev \& Kovalev(2012)]{r:pus12}
      Pushkarev~A.~B., Kovalev~Y.~Y., 2012, \aa, 544, 34

   \bibitem[Sault, Teuben \& Wright(1995)]{r:miriad}
      Sault~R.~J., Teuben~P.~J., Wright~M.~C.~H., 1995, In Astronomical Data 
        Analysis Software and Systems IV, ed. by R.~Shaw, H.E.~Payne, 
        and~J.J.E. Hayes, ASP Conference Series, 77, 433

   \bibitem[Taylor et al.(2007)]{r:tay07}
      Taylor~G.~B., Healey~S.~E., Helmboldt~J.~F., et al.\
         2007, \apj, 671, 1355

   \bibitem[Tsarevsky et al.(2005)]{r:tsa}
      Tsarevsky~G. et al.,
	2005, \aa, 438, 949

   \bibitem[Wieringa \& Kesteven (1992)]{r:atca_beam}
        Wieringa~M.~H., Kesteven~M.~J.\
           1992, ATNF Technical Memo, 39.2/010 \\
           \web{http://www.atnf.csiro.au/observers/memos/d96b7e~1.pdf}

   \bibitem[Wilson et al.(2011)]{r:cabb}
      Wilson~W.~E. et al.,
      2011, \mnras, 416, 832

   \bibitem[Wright et~al.(2010)]{r:wise} 
      Wright~E.~L. et al.,
         2010, \aj, 140, 1868

\end{thebibliography}
\end{document}